\documentclass[twocolumn,amsmath,amssymb]{revtex4}
\usepackage{graphicx}
\usepackage{dcolumn}
\usepackage{bm}
\begin{document}

\title{Tilt-Induced Phase Transitions in Even-Denominator Fractional Quantum Hall
States at the ZnO Interface}
\author{Wenchen Luo and Tapash Chakraborty\footnote{Tapash.Chakraborty@umanitoba.ca}}
\affiliation{Department of Physics and Astronomy, University of Manitoba, Winnipeg,
Canada R3T 2N2}

\date{\today}
\begin{abstract}
Even denominator fractional quantum Hall states in a ZnO quantum well reveal interesting phase 
transitions in a tilted magnetic field. We have analyzed the planar electron gas in ZnO, 
confined in a parabolic potential in the third dimension, perpendicular to the plane of the 
electron gas. Since the Landau level gap is very small in this system we have included the 
screened Coulomb potential in order to include the effects of all the Landau levels. We observe 
an incompressible state - compressible state phase transition induced by the tilted field. 
Additionally, the 5/2 state has been experimentally found to be missing in this system. We however propose 
that a wider quantum well may help to stabilize the incompressible phase at the 5/2 filling factor.
\end{abstract}

\maketitle
The `enigmatic' even-denominator fractional quantum Hall effect (FQHE) at the filling factor $\nu=\frac52$
was first discovered in GaAs heterojunctions \cite{eisenstein,willett}. It is a special member in the
FQHE family since its ground state and excitations can not be described by the Laughlin wave 
function \cite{laughlin,book}. Although there are some aspects of this state that still remains
unclear as yet, it is generally believed that a Pfaffian state with non-Abelian excitations is the
most likely candidate to describe this extrodinary FQHE \cite{Read,Bilayer2}. Numerical studies 
of the even denominator FQHE were also helpful to understand the nature of this state 
\cite{TC_even,peterson2,rezayi,papic}. The ground state at this filling factor is incompressible (just as for 
the odd-denominator states), so that the attributes of the transport experiments are the same as those of 
the odd-denominator FQHE. Recently, the FQHE has been discovered in an oxide material, the ZnO interface 
with high miobility \cite{zno,tsukazaki,falson}. There is heightened expectations that the stronger Coulomb 
interactions in this system will perhaps display unsual effects related to strong electron correlations 
\cite{mannhart}. Interestingly, in the MgZnO/ZnO interface the $\nu=\frac52$ state was found to be missing. 
The ZnO quantum well is quite different from the conventional GaAs system because here the Landau level (LL) 
gap is comparable to the Zeeman gap and the ratio $\kappa$ of the Coulomb interaction to the LL gap is very large. 
In order to explain the missing FQHE, we earlier introduced the screened Coulomb interaction which includes the 
effect of all the LLs in our model \cite{luo3}. The system was indeed found to be compressible, thereby 
explaining the absence of the $\nu=\frac52$ state in ZnO.

A tilted magnetic field is a very useful means to study the nature of the fractional quantum Hall 
systems \cite{jim_tilted,dean,xia}. It modifies the transport properties to provide additional information 
about the systems, especially the spin polarization and the excitations asscoiated with electron spins 
\cite{FQHE_spin,tapash2}. In the experiment by Falson et al. involving the ZnO \cite{falson}, the tilted magnetic 
field reflects very unusual behavior in the transport properties. At the $\nu=\frac32$ filling
factor the FQHE was found to appear only when the tilt angle is large. On the other hand, for the $\nu=\frac72$ filling
factor the FQHE disappeared when the tilt angle was increased. We believe that there are phase transitions 
associated with these experiemtal observations; since the LL energy gap which is only related to the perpendicular 
component of the magnetic field can be easily exceeded by the Zeeman coupling which is propotional to the
total magnetic field. In particular, the spin transition is also involved in these phase transitions. 
The spin polarization measurements could therefore be used to observe these phase transitions. With the increase 
of the tilt angle the screening potential is changed. So the transport properties of the electron gas 
in a variable tilted field must become very different from those in a perpendicular magnetic field alone. 
Here we focus only on the even-denominator filling factors.

In our work, the electron gas is confined in a parabolic potential in the $z$ direction, thereby making 
the system quasi-two-dimensional. Motivated by the experimental work of Falson et al. \cite{falson}, we 
consider this system in a tilted magnetic field. In the ZnO quantum well, unlike in the GaAs system, the 
LL crossing is easily achieved in a tilted field. Since the Zeeman coupling is about $0.94$ of the LL 
gap in a perpendicular magnetic field, a spin polarized state may not satisfactorily describe the system. We 
therefore use a spin-mixed Hamiltonian to study the ground states and the phase transitions involving 
spin flipping.

In our previous work \cite{luo3} we proposed that the missing fractional quantum Hall state at 
$\nu=\frac52$ is due to the screened Coulomb potential. We also use that screened Coulomb potential to 
study the FQHE in a tilted magnetic field. With a non-zero tilt angle $\alpha^{}_0$, it is more 
realistic to suppose that the electron gas is confined in a parabolic potential with the frequency 
$\omega^{}_z$ in the $z$ direction which is perpendicular to the plane of the electron gas. The 
advantage of this approximate potential is that the wave function can be analytically obtained 
for any value of the tilted field. This approximation to a certain extent should be similar to 
other finite-thickness approaches such as the infinite square well \cite{peterson2} or a triangular 
well. We choose the Landau gauge with the vector potential $\mathbf{A}=\left(0,B^{}_zx-B^{}_xz,0\right).$ 
The wave function is then given by \cite{papic,wang,yang,tilted} 
\begin{eqnarray*}
\psi^{}_{k,n^{}_1,n^{}_2}\left( \mathbf{u}\right) &=&\frac{e^{iky}}{\sqrt{L^{}_y}}
\phi_{n^{}_1}^1\left[ \left( x+k\ell^2\right)\cos\theta-z\sin\theta 
\right]  \nonumber \\
&&\times \phi_{n_2}^2\left[\left(x+k\ell^2\right)\sin\theta+z\cos\theta \right],
\end{eqnarray*}
where $k$ is the guiding center index, $n^{}_{1,2}$ is the sub LL index, and
\begin{equation*}
\theta =\arctan \frac{\omega_b^2-\omega_{\perp}^2+\sqrt{\left(
\omega_b^2-\omega_{\perp}^2\right)^2+4\omega_{\parallel}^2\omega_{\perp}^2}}{2
\omega^{}_{\parallel}\omega^{}_{\perp}}
\end{equation*}
with $\omega^{}_{\perp}=eB^{}_z/m^{\ast}$, $\omega^{}_{\parallel}=eB^{}_x/m^{\ast}$ and 
$\omega_b^2=\omega_{\parallel}^2+\omega_z^2.$ The wave function of a harmonic 
oscillator is 
$\phi_n^i\left(x\right)=\frac1{\sqrt{\sqrt{\pi}2^nn!\ell^{}_i}}
\exp \left( -\frac{x^2}{2\ell_i^2}\right) H^{}_n\left(\frac{x}{\ell^{}_i}\right),$
with a Hermite polynomial $H^{}_n$. The magnetic length is $\ell =\sqrt{c/eB^{}_z}$ and 
the magnetic length for each subband is $\ell^{}_{1,2}=1/\sqrt{m^{\ast}\omega^{}_{1,2}},$ 
where $m^{\ast}$ is the effective mass and $\omega^{}_{1,2}=\frac1{\sqrt{2}}\sqrt{
\omega_b^2+\omega_{\perp}^2\pm \sqrt{\left(\omega_b^2-\omega_{\perp}^2\right)^2+
4\omega_{\parallel}^2\omega_{\perp}^2}}.$ The energies of the LLs are 
$E^{}_{n^{}_1n^{}_2}=\left(n^{}_1+\frac12\right)\hbar\omega^{}_1+\left(n^{}_2+\frac12\right) 
\hbar\omega^{}_2,$ where $n^{}_1,n^{}_2$ are sub LL indices. In a parabolic potential the LL indices must be 
described by two branches. One is related to the original LL in a pure two-dimensional 
(2D) case, while the other one is essentially the energy levels of the parabolic potential.

The interaction Hamiltonian including spin is given by
\begin{eqnarray*}
&&H=\frac12\sum^{}_{\alpha,\beta}\sum^{}_{m^{\left(\prime\right)},n^{\left(\prime\right) 
}}\sum^{}_{j^{}_1\ldots j^{}_4}\int d\mathbf{u}^{}_1 d\mathbf{u}^{}_2\psi_{j^{}_1,m,n}^{\ast}\left( 
\mathbf{u}^{}_1\right) 
\nonumber \\
&&\psi_{j^{}_2,m^{\prime},n^{\prime}}^{\ast}\left(\mathbf{u}^{}_2\right)
V\left( \mathbf{u}^{}_1\mathbf{-u}^{}_2\right) \psi^{}_{j^{}_3,m^{\prime},n^{\prime}}\left( 
\mathbf{u}^{}_2\right) \psi^{}_{j^{}_4,m,n}\left( \mathbf{u}^{}_1\right)  \nonumber \\
&&c_{\alpha,j^{}_1,m,n}^{\dag}c_{\beta,j^{}_2,m^{\prime},n^{\prime}}^{\dag}c^{}_{\beta,j^{}_3,
m^{\prime},n^{\prime}}c^{}_{\alpha,j^{}_4,m,n},
\end{eqnarray*}
where we only consider one LL or two LLs with different spins. In the
rectangular space, the discret momenta is $q^{}_{x}=\frac{2\pi }{L^{}_{x}}s,q^{}_{y}=
\frac{2\pi }{L^{}_{y}}t$. So the Coulomb interaction is then given by \cite{tilted,halonen}
\begin{eqnarray*}
&&V^{}_{C}=\frac{2e^2}{\epsilon \ell}\frac1{S}\overline{\sum^{}_{q^{}_x,q^{}_y}}
\delta_{j^{}_1,j^{}_4+q^{}_y}^{\prime}\delta_{j^{}_2,j^{}_3-q^{}_y}^{\prime}e^{i\left(j^{}_3-j^{}_1
\right) q^{}_x\ell^2-q_1^2-q_2^2}  \nonumber
\\
&&\int \frac{dq^{}_z\ell}{\epsilon^{}_s\left(\mathbf{q}\right)}\exp\left(-q_{-}^2-q_{+}^2
\right) \frac{\Xi\left(m,n;\mathbf{q}\right)\Xi\left( m^{\prime},n^{\prime};\mathbf{q}\right)}{
q_x^2+q_y^2+q_z^2},
\end{eqnarray*}
where $\epsilon$ is the dielectric constant, $S=L^{}_xL^{}_y$ is the area of
the sample, $\delta^{\prime}$ includes the periodic boundary condition and
the bar over the summation excludes the term $q^{}_x=q^{}_y=0$. We also write
\begin{eqnarray*}
q^{}_1 &=&\frac1{\sqrt2}\frac{\cos\theta}{\ell^{}_1}q^{}_y\ell^2, \qquad 
q^{}_2 = \frac1{\sqrt2}\frac{\sin\theta}{\ell^{}_2}q^{}_y\ell^2 \\
q^{}_- &=&\frac1{\sqrt2}\left(q^{}_z\sin\theta-q^{}_x\cos\theta\right)\ell^{}_1, \\
q^{}_+ &=& \frac1{\sqrt2}\left(q^{}_z\cos\theta+q^{}_x\sin\theta \right) \ell^{}_2 \\
\lambda^{}_{1,2}\left(m,m^{\prime},\mathbf{q}\right) &=&\left[\mathtt{sign}
\left( m-m^{\prime}\right) q^{}_{1,2}\mp iq^{}_{\mp}\right]^{\left\vert
m-m^{\prime}\right\vert}  \nonumber \\
&&L_{\min\left(m,m^{\prime}\right)}^{\left\vert m-m^{\prime}\right\vert}
\left(q_{1,2}^2+q_{\mp}^2\right)
\end{eqnarray*}
with a Laguerre polynomial $L$, and
\begin{equation*}
\Xi \left( m,n;\mathbf{q}\right)=\lambda^{}_1\left( m,m,\mathbf{q}\right)
\lambda^{}_2\left( n,n,\mathbf{q}\right).
\end{equation*}
Following Ref.~\cite{luo3}, we introduce the dielectric function of the
screening in a general three-dimensional case, 
\begin{equation*}
\epsilon^{}_s\left(\mathbf{q}\right)=1-V\left(\mathbf{q}\right) \chi_{nn}^0
\left(\mathbf{q}\right)=1-\frac{2\pi e^2}{q^2\epsilon}\chi_{nn}^0\left( 
\mathbf{q}\right).
\end{equation*}
The non-interacting density-density response function in the Matsubara frequency 
$\Omega^{}_n$ is 
\begin{eqnarray*}
\chi_{nn}^0\left(\mathbf{q,} i\Omega^{}_n \right) &=&\frac{N^{}_s}{SL^{}_z}
\sum_{\sigma,m,m^{\prime},n,n^{\prime}}^{{}}\left\vert G^{}_{mn,m^{\prime}
n^{\prime}}\left(-\mathbf{q}\right) \right\vert^2  \nonumber \\
&&\times \frac{\nu_{\sigma,mn}^{{}}-\nu_{\sigma,m^{\prime}n^{\prime
}}^{{}}}{i\hbar \Omega_n^{{}}+\left( E_{mn}^{{}}-E_{m^{\prime }n^{\prime
}}^{{}}\right)},
\end{eqnarray*}
where $N^{}_s$ is the degeneracy of a LL, $\sigma$ is the spin index, $m^{\left( 
\prime\right)},n^{\left(\prime\right)}$ are the sub LL indices, $\nu$ is the 
filling factor, $E_{mn}^{{}}$ is the kinetic energy of the sub LL $\left( 
m,n\right)$, and the form factor is defined as 
\begin{eqnarray*}
&&G_{mn,m^{\prime}n^{\prime}}\left(\mathbf{q}\right) =\exp \left[-\frac12
\left(q_1^2+q_2^2+q_-^2+q_+^2\right)\right] \\
&&\sqrt{\frac{\min\left(m,m^{\prime}\right)!\min\left(n,n^{\prime}\right)!}{
\max \left(m,m^{\prime}\right)!\max \left( n,n^{\prime}\right)!}}\lambda_1
\left(m,m^{\prime},\mathbf{q}\right) \lambda_2\left(n,n^{\prime},\mathbf{q}\right).  
\nonumber
\end{eqnarray*}
For the third dimension, the length in the $z$ direction, $L^{}_z$ is difficult to determine 
in a parabolic potential. In principle it should be infinity, but for our present purpose 
we choose a finite value for $L^{}_z$ since the wave functions vanish rapidly in the $z$ direction. 
The electrons are confined well in the $z$ direction, so it is reasonable to limit $L^{}_{z}$ 
in a proper range. Here, we approximate the $L^{}_z$ in terms of the parabolic potential frequency 
$\omega^{}_z$, $L^{}_z=2\sqrt{\ln\left(2\right)\frac{\omega^{}_{\bot}}{\omega^{}_z}}\ell,$
which is the width of the wave function of the lowest LL in the $z$ direction. We 
set $L^{}_z=1.8$ nm which corresponds to a ralatively narrow quantum well, since the electron gas 
will be split into a "double" layer system in a wide quantum well \cite{shabani} and the
consequent transport properties would become very different. The density-density response function 
is calculated in the noninteracting case, so the filling factors $\nu$ are also the noninteracting 
filling factors.

We utilize the exact diagnoalization scheme in the standard case of finite-size systems in a 
periodic rectangular geometry $\left(L^{}_x=L^{}_y\right)$ \cite{book,haldane} with the screening 
potential included \cite{luo3,shizuya}. In order to analyze the phase transition near the LL crossing 
in a tilted field, we take one LL or two LLs with different spin orientations. With an increase 
of the tilt angle, the Landau level crossings occur. The first crossing 
happens at about $18^{\circ}$, while the secend one at about $62^{\circ}$. We study only the small 
tilt angles, since the parabolic potential may not be a very good approximation at very large 
angle. When the quantum well is narrow, the energies of the series of sub LLs $(m>0,0)$ are much 
higher than those of sub LLs $\left(0,n<3\right)$. When the tilt angle is $\alpha^{}_0<60^{\circ}$, 
the LL crossing only happens between the LLs $\left(0,n;\downarrow\right)$ and $\left(0,n+1;\uparrow\right)$. 
The magnetic fields of the $z$-component are set the same as in Ref.~\cite{falson}, i.e.,
$B^{}_z=6.2,3.75,2.75$ T for $3/2,5/2,7/2$, respectively.

\textit{Phase transition at }$\nu=\frac32$\textit{:} We consider the two LLs 
$\left( 0,0;\downarrow \right) $ and $\left( 0,1;\uparrow \right) $ into our exact diagonalization 
scheme. We did not find any spin coherence irrespective of the tilt angle. There is a phase transition
associated with spin polarization at about $\alpha ^{}_{1}=23^{\circ }$. When $\alpha ^{}_{0} <
\alpha ^{}_{1}$, all electrons are in LL $\left( 0,0;\downarrow \right)$, even though the kinetic energy 
of $\left( 0,1;\uparrow \right)$ is lower when $18^{\circ }<\alpha^{}_0 <\alpha ^{}_{1}$. The incompressiblity 
of the ground state may not be stable for different sizes of the system. When $\alpha^{}_{0}>\alpha ^{}_{1}$, 
all electrons flip to $\left( 0,1;\uparrow \right)$. In this LL, the even-denominator FQHE can be 
found as an incompressible liquid without any LL mixing or screening. For $\nu=\frac32$, the screening is weaker 
than that of filling factor $\nu=\frac52$. The collective modes for $\nu=\frac32$ when $\alpha^{}_{0} >
\alpha^{}_{1}$ show that the ground state is incompressible. However, in the experiment, the FQHE 
is only observed when $\alpha^{}_{0} > 38^{\circ}$ \cite{falson}.

\begin{figure}[tbp]
\includegraphics[width=7.0cm]{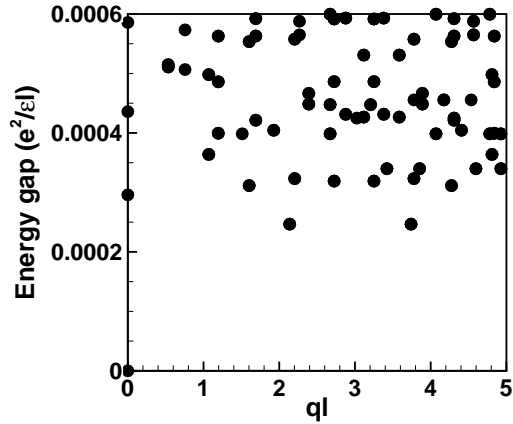}
\caption{The collective mode of $\protect\nu =7/2$ for $11$ electrons in a
tilted magnetic field ($\theta=10^\circ$). }
\label{figure1}
\end{figure}

\textit{Phase transitions at} $\nu=\frac72$\textit{:} When the 
magnetic field is perpendicular to the electron plane, we have already shown that the FQHE survives the 
screening potential \cite{luo3}. In a tilted field, with the parabolic potential we consider the two LLs 
$\left(0,1;\downarrow \right)$ and $\left( 0,2;\uparrow \right)$ into the exact diagonalization scheme. 
The phase transition which is found to be at $\alpha^{}_2=18^{\circ}$ is more or less 
at the same angle where the non-interacting LLs cross. The incompressible ground state is still found
when $\alpha ^{}_0<\alpha ^{}_2$. All electrons are in LL $\left(0,1;\downarrow \right)$, 
and the collective mode for eleven electrons is shown in Fig.~\ref{figure1}. 

The collective modes do not change much when $\alpha^{}_0 < \alpha^{}_2$. The two minima are at about 
$q\ell =2.2$ and $3.8$, which is close to the collective mode of the pure 2D case. When $\alpha^{}_0 >
\alpha^{}_2$, the screening is changed due to the fact that the non-interacting filling factors are changed,
and the electrons are flipped to LL $\left(0,2;\uparrow\right)$, and the system is no longer incompressible. 
In the experiment, the phase transition occurs in the range $\left(21^{\circ},27^{\circ}\right)$ \cite{falson}. 
The difference between our theoretical work and the measurement is likely due to the LL broadening 
which is induced by the disorder. The LL broadening is able to shift the LL crossing to a higher tilt angle.
The second LL crossing occurs for $\alpha^{}_0=62^{\circ}$, between $\left(0,2;\uparrow\right)$ and 
$\left(0,0;\downarrow \right)$. We determine the collective modes of LL $\left(0,0;\downarrow\right)$ and
the ground state is still compressible. In the phase transitions at this filling factor, the 
mixed-spin state is also absent. No spin coherence state has lower energy than the spin polarized state. 
All phase transitions are first-order.

\begin{figure}[tbp]
\includegraphics[width=6.0cm]{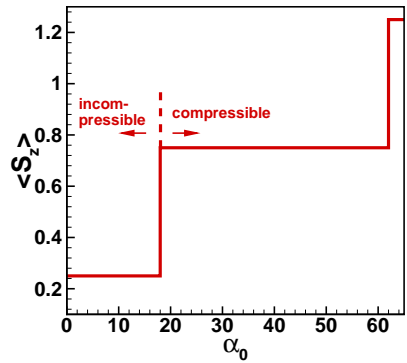}
\caption{The phase transitions at the $\nu=\frac72$ filling factor.}
\label{figure2}
\end{figure}

We note here that all phase transitions involve spin flip. The spin polariztion before and after 
all the phase transitions is changed significantly. If we define the spin polarization as $\left\langle
S^{}_z\right\rangle =\frac12\left(\nu^{}_{\uparrow}-\nu^{}_{\downarrow}\right)$, then the spin phase 
transition is shown in Fig.~\ref{figure2}. Therefore, the spin polarization measurement could be a very 
powerful means to determine the phase transitons. We expect that future experimental work may confirm our 
present findings.

\begin{figure}[tbp]
\includegraphics[width=7.0cm]{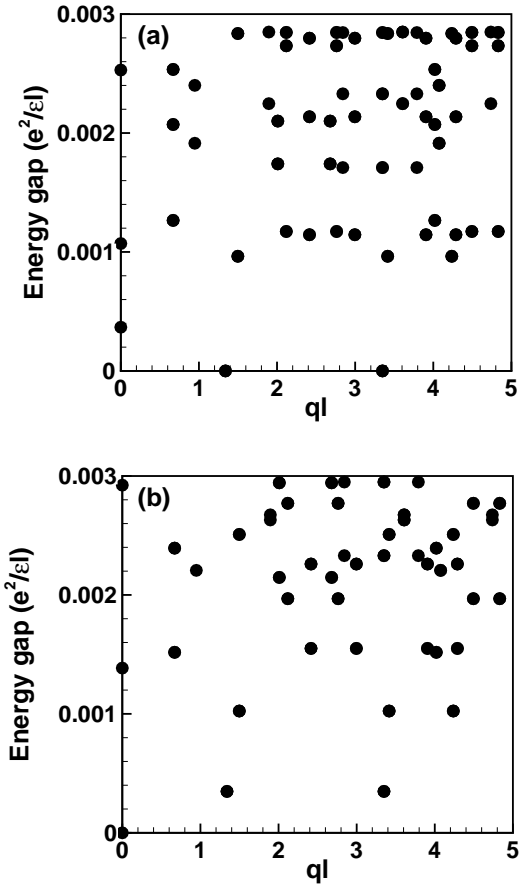}
\caption{The collective mode of $\protect\nu =\nu=\frac52$ for seven electrons in a
perpendicular magnetic field. (a) The characterised width of the parabolic
potential $L^{}_{z}=1.8$ nm and (b) $L^{}_{z}=11$ nm.}
\label{figure3}
\end{figure}

At filling factor $\nu=\frac52$, in a pure 2D case the FQHE is absent due to the screened Coulomb potential
from other LLs \cite{luo3}. The FQHE is still absent for the parabolic potential. Interestingly, the $\nu=\frac52$
FQHE could occur if we tune the width of the quantum well. When $L^{}_z\geq 9$ nm, the ground state of $\nu=\frac52$
becomes incompressible. Figure~\ref{figure3} (a) and Fig.~\ref{figure3} (b) display the collective modes of 
seven electrons for $L^{}_z=1.8$ nm and $11$ nm in a zero tilted field, respectively. When the quantum well 
is narrow the electron gas behaves similar to a pure 2D case. But the collective modes show that the 
incompressibility appears when the width is increased. However, in a higher magnetic field a narrower width 
can also stabilize the FQHE state. In GaAs, both theoretical and experimental works indicated that wider quantum well 
helps to stablize the FQHE at $\nu=\frac52$ \cite{peterson2,xia}. Our results are therefore compatible 
with those previous works. However, we also find that the other even-denominator state at $\nu=\frac72$ would become 
compressible when the width of quantum well is increased. If the width of the quantum well in ZnO could be 
artifically fabricated then we hope that this width effect predicted here can indeed be observed.

In order to explore the nature of the ground state, we have calculated the wave functions 
of the even-denominator FQHE states. For simplicity, we consider the pure 2D case. The many-body wave 
function of the ground state at $\nu=\frac72$ in a perpendicular magnetic field is $\left\vert \phi_{\frac72}^s
\right\rangle$ with our screening potential. The wave function of $\nu=\frac52$ or $\nu=\frac72$ without screening is 
$\left\vert \phi^{}_{\frac52}\right\rangle =\left\vert \phi^{}_{\frac72}\right\rangle$. We find that the overlap 
of these two wave functions is close to unity, $\left\langle \phi^{}_{\frac72}|\phi_{\frac72}^{s}\right\rangle
\approx 0.99$. So the screened Coulomb potential at $\nu=\frac72$ does not change the wave function. The 
pair distribution function is also similar to that of a liquid \cite{TC_even}. Note that the screened
Coulomb potential at $\nu=\frac52$ completely destroy the liquid ground state, the lowest energy state is a 
density wave.

In summary, we have analyzed the phase transitions at $\nu=\frac32$ and $\nu=\frac72$ in the exact diagonalization 
scheme of the two LLs with different spin polarization when the magnetic field is tilted from the direction
perpendicular to the electron plane. We find phase transitions between different spins. The 
nature of the ground states is changed significantly in the phase transition. However, we could not find 
any spin coherent state in the phase transition. It seems that the spin polarization is first-order-like 
flipped to the other spin orientation when the tilt angle is increased. The spin flipping 
could be observed in a spin-sensitive experiment \cite{muraki}. Experimentally, the $\nu=\frac52$ FQHE was found
to be absent in ZnO. However, we propose that if the width of the quantum well could be artificially
widened then this FQHE state might be observable. Incidentally, the measured activation gaps in ZnO are usually one 
order of magnitude smaller than those in GaAs \cite{tsukazaki}. This can be qualitatively understood
as follows: the LL gap of GaAs is about seven times larger than that of ZnO, and therefore the screening 
of ZnO is seven times stronger than that of the GaAs. Consequently, the excitation energy in ZnO is about 
one order smaller than that in GaAs.

The work has been supported by the Canada Research Chairs Program of the
Government of Canada. The computation time was provied by Calcul Qu\'{e}bec
and Compute Canada.


\begin{thebibliography}{99}
\bibitem{eisenstein} J.P. Eisenstein, in {\it Perspectives in Quantum Hall
Effects,} edited by S. Das Sarma and A. Pinczuk (Wiley-Interscience, New York, 
1996), p. 37.

\bibitem{willett} R. Willett, J.P. Eisenstein, H.L. St\"ormer, D.C. Tsui,
A.C. Gossard, and J.E. English, Phys. Rev. Lett. \textbf{59}, 1776 (1987).

\bibitem{laughlin} R. B. Laughlin, Phys. Rev. Lett. \textbf{50}, 1395 (1983).

\bibitem{book} T. Chakraborty and P. Pietil\"ainen, \textit{The Quantum Hall
Effects} (Springer, New York, 1995); \textit{The Fractional Quantum Hall
Effect} (Springer, New York, 1988).

\bibitem{Read} N. Read, Physica \textbf{B 298}, 121 (2001); G. Moore and N.
Read, Nucl. Phys. B \textbf{360}, 362 (1991).

\bibitem{Bilayer2} V.M. Apalkov and T. Chakraborty, Phys. Rev. Lett. \textbf{%
107}, 186803 (2011).

\bibitem{TC_even} T. Chakraborty and P. Pietil\"{a}inen, Phys. Rev. B Phys.
Rev. B \textbf{38}, 10 097(R) (1988).

\bibitem{peterson2} M. R. Peterson, Th. Jolicoeur, and S. Das Sarma, Phys.
Rev. Lett. \textbf{101}, 016807 (2008).

\bibitem{rezayi} E.H. Rezayi and S.H. Simon, Phys. Rev. Lett. \textbf{106},
116801 (2011), M.R. Peterson and C. Nayak, Phys. Rev. Lett. \textbf{113},
086401 (2014).

\bibitem{papic} Z. Papi\'{c}, Phys. Rev. B \textbf{87}, 245315 (2013).

\bibitem{zno} A. Tsukazaki1, A. Ohtomo, T. Kita, Y. Ohno, H. Ohno, and M.
Kawasaki, Science \textbf{315}, 1388 (2007).

\bibitem{tsukazaki} A. Tsukazaki, S. Akasaka, K. Nakahara, Y. Ohno, H. Ohno,
D. Maryenko, A. Ohtomo and M. Kawasaki, Nature Materials \textbf{9}, 889
(2010).

\bibitem{falson} J. Falson, D. Maryenko, B. Friess, D. Zhang, Y. Kozuka, A.
Tsukazaki, J.H. Smet and M. Kawasaki, Nature Physics \textbf{11}, 347 (2015).

\bibitem{mannhart}
J. Mannhart and D.G. Schlom, Science {\bf 327}, 1607 (2010); J. Mannhart, D.H.A. Blank,
H.Y. Hwang, A.J. Millis, and J.-M. Triscone, MRS Bulletin {\bf 33}, 1027 (2008);
H.Y. Hwang, Y. Iwasa, M. Kawasaki, B. Keimer, N. Nagaosa, and Y. Tokura, Nat. Mater.
{\bf 11}, 103 (2012); Y. Kozuka, A. Tsukazaki, and M. Kawasaki,
Appl. Phys. Rev. {\bf 1}, 011303 (2014).

\bibitem{luo3} W. Luo and T. Chakraborty, Phys. Rev. B \textbf{93},
161103(R) (2016).

\bibitem{jim_tilted} J.P. Eisenstein, R. Willett, H.L. St\"ormer, D.C. Tsui,
A.C. Gossard, and J.E. English, Phys. Rev. Lett. \textbf{61}, 997 (1988).

\bibitem{dean} C. R. Dean, B. A. Piot, P. Hayden, S. Das Sarma, G. Gervais,
L. N. Pfeiffer, and K. W. West, Phys. Rev. Lett. \textbf{101}, 186806 (2008).

\bibitem{xia} J. Xia, V. Cvicek, J. P. Eisenstein, L. N. Pfeiffer, and K. W.
West, Phys. Rev. Lett. \textbf{105}, 176807 (2010).

\bibitem{FQHE_spin} T. Chakraborty, P. Pietil\"ainen, and F.C. Zhang, Phys.
Rev. Lett. \textbf{57}, 130 (1986); J.P. Eisenstein, H.L. Stormer, L.
Pfeiffer, and K.W. West, Phys. Rev. Lett. \textbf{62}, 1540 (1989).

\bibitem{tapash2} T. Chakraborty and F.C. Zhang Phys. Rev. B \textbf{29},
7032(R) (1984); F.C. Zhang and T. Chakraborty, Phys. Rev. B \textbf{30},
7320(R) (1984).

\bibitem{tilted} T. Chakraborty and P. Pietil\"ainen, Phys. Rev. B {\bf 39},
7971 (1989); V. Halonen, P. Pietil\"ainen and T. Chakraborty, Phys. Rev. B
{\bf 41}, 10202 (1990). 

\bibitem{wang} Daw-Wei Wang, S. Das Sarma, Eugene Demler, and Bertrand I.
Halperin, Phys. Rev. B \textbf{66}, 195334 (2002).

\bibitem{yang} Bo Yang, Z. Papi\'{c}, E. H. Rezayi, R. N. Bhatt, and F. D.
M. Haldane, Phys. Rev. B \textbf{85}, 165318 (2012).

\bibitem{halonen} V. Halonen, Phys. Rev. B \textbf{47}, 4003 (1993), Phys.
Rev. B \textbf{47}, 10001(R) (1993).

\bibitem{shabani} J. Shabani, Yang Liu, M. Shayegan, L. N. Pfeiffer, K. W.
West, and K. W. Baldwin, Phys. Rev. B \textbf{88}, 245413 (2013).

\bibitem{yoshioka3} D. Yoshioka, J. Phys. Soc. Jpn. \textbf{55}, 3960-3968
(1986); M. Rasolt, F. Perrot, and A.H. MacDonald, Phys. Rev. Lett. {\bf 55}, 
433 (1985).

\bibitem{haldane} F.D.M. Haldane, Phys. Rev. Lett. \textbf{55}, 2095 (1985).

\bibitem{shizuya} K. Shizuya, Phys. Rev. B \textbf{75}, 245417 (2007); R.
Roldan, M.O. Goerbig, and J.-N. Fuchs, Semicond. Sci. Technol. \textbf{25},
034005 (2010). W. Luo and R. C\^{o}t\'{e}, Phys. Rev. B \textbf{88}, 115417
(2013).


\bibitem{muraki} L. Tiemann, G. Gamez, N. Kumada, and M. Muraki, Science 
\textbf{335}, 828 (2012); M. Stern, B. A. Piot, Y. Vardi, V. Umansky, P.
Plochocka, D. K. Maude, and I. Bar-Joseph, Phys. Rev. Lett. \textbf{108},
066810 (2012).

\end{thebibliography}
\end{document}